\documentclass{article}

\usepackage{arxiv}

\usepackage[utf8]{inputenc} % allow utf-8 input
\usepackage[T1]{fontenc}    % use 8-bit T1 fonts
\usepackage{hyperref}       % hyperlinks
\usepackage{url}            % simple URL typesetting
\usepackage{booktabs}       % professional-quality tables
\usepackage{amsfonts}       % blackboard math symbols
\usepackage{nicefrac}       % compact symbols for 1/2, etc.
\usepackage{microtype}      % microtypography
\usepackage{lipsum}

\usepackage[citestyle=authoryear,natbib=true]{biblatex}
\usepackage{graphicx}
\usepackage{subcaption}
\usepackage{color}
\usepackage{wrapfig}

\title{Inferring Astrophysical X-Ray Polarization with Deep Learning}

\author{
 Nikita Moriakov \\
  Radiology, Nuclear Medicine and Anatomy \\
  Radboud University Medical Center \\
  Nijmegen, Netherlands \\
  \texttt{nikita.moriakov@radboudumc.nl} \\
   \And
 Ashwin Samudre \\
  Aix-Marseille-Univ, CNRS, CINAM\\
  Marseille, France\\
  \texttt{ashwin.samudre@cinam.univ-mrs.fr} \\
   \And
 Michela Negro \\
  NASA-GSFC/CRESST-UMBC\\
  Greenbelt, MD 20771, USA\\
  \texttt{michela.negro@nasa.gov} \\
   \And
 Fabian Gieseke \\
  Department of Computer Science\\
  University of Copenhagen\\
  Copenhagen, Denmark \\
  \texttt{fabian.gieseke@di.ku.dk} \\
   \And
 Sydney Otten \\
  University of Amsterdam and \\
  Radboud University Nijmegen \\
  \texttt{s.m.m.otten@uva.nl}
   \And
 Luc Hendriks \\
  Radboud University \\
  Nijmegen, the Netherlands \\
  \texttt{luc@luchendriks.com}\\
}

\date{}

\begin{document}
\maketitle
\begin{center}
    \textbf{DarkMachines collaboration: \url{darkmachines.org}. Particle Track Reconstruction with ML Group.}\\
\end{center}

\begin{abstract}
We investigate the use of deep learning in the context of X-ray polarization detection from astrophysical sources as will be observed by the Imaging X-ray Polarimetry Explorer (IXPE), a future NASA selected space-based mission expected to be operative in 2021. In particular, we propose two models that can be used to estimate the impact point as well as the polarization direction of the incoming radiation. The results obtained show that data-driven approaches depict a promising alternative to the existing analytical approaches. We also discuss problems and challenges to be addressed in the near future.
% This works addresses an important problem in astrophysics. We propose the IXPE mission and the associated data. We investigate the use of modern deep learning techniques and provide the experimental outcome, which shows that the data-driven approach depicts a promising competitor to the the analytical approach commonly employed in this field. We also sketch a number of promising potential improvements to our deep learning baselines.
\end{abstract}

% keywords can be removed

\section{Introduction}

The 
\emph{Imaging X-ray Polarimetry Explorer}\footnote{\url{https://ixpe.msfc.nasa.gov}} (IXPE), which is a space-based mission selected by the NASA, is expected to be launched in 2021. IXPE will conduct precise polarimetry in the X-ray energy band (between 2 and 8 kilo electronvolts), which is a poorly investigated field so far, see \citep{galaxies6010033} for a recent review. The data collected by this mission will be important for the analysis of various 
%In this work , we will investigate several different categories of 
astrophysical sources, from stellar-mass black holes, neutron stars and pulsar wind nebulae, to supernovae remnants and active galactic nuclei. 

\begin{wrapfigure}{r}{8.0cm}
    \vspace{-8pt}
    \centering
    %\hskip{-10pt}
\includegraphics[width=7.5cm]{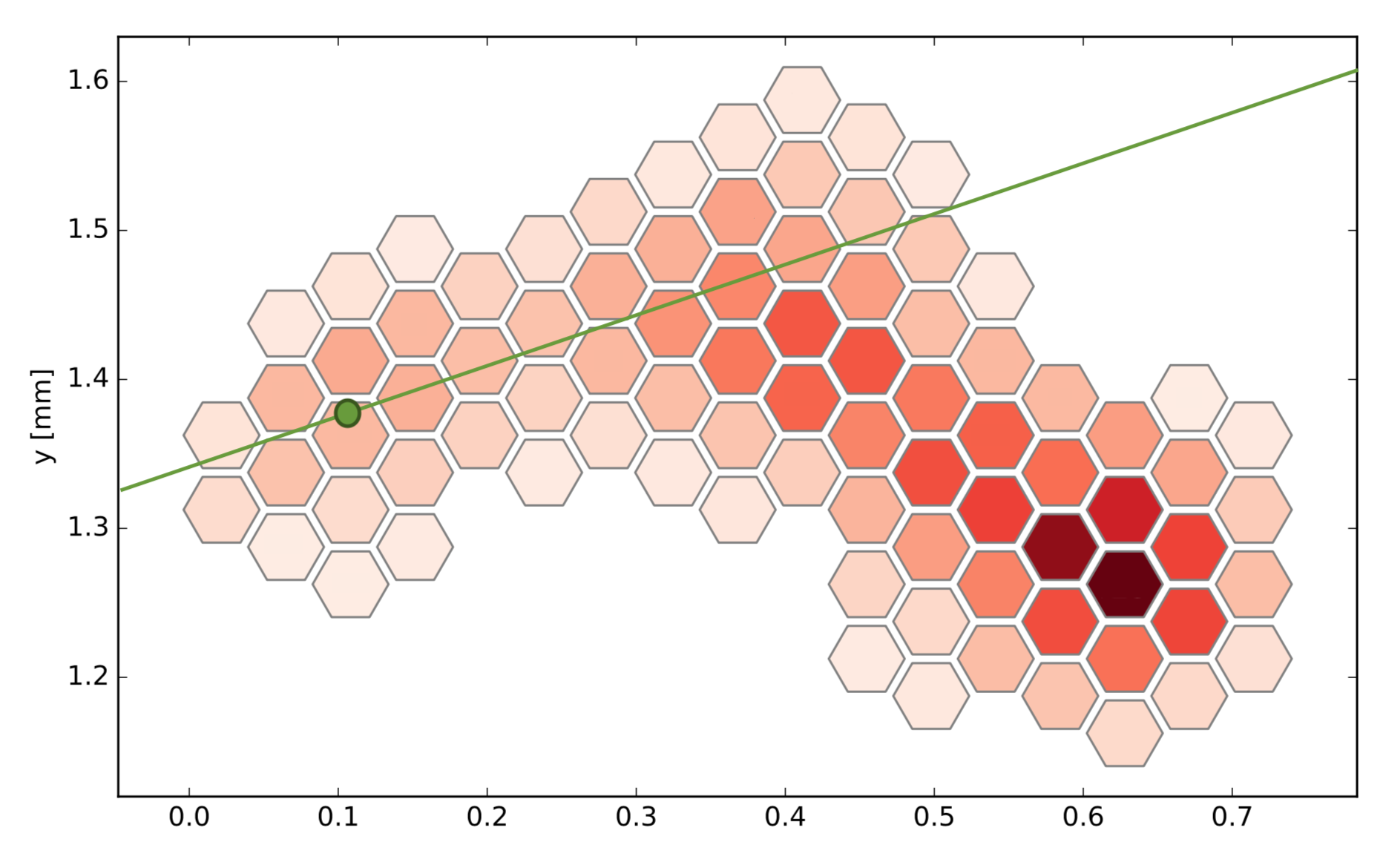}
    %\vspace{-5pt}
    \caption{Example of a PE track: % at 6 keV in IXPE detector. 
    The darker the color, the higher the charge value recorded.}
    \label{fig:ixpetrack}
    \vspace{-8pt}
\end{wrapfigure}

IXPE exploits the so-called \emph{Gas Pixel Detector}~(GPD) design to perform measurements of linear polarization~\citep{2006NIMPA.560..425B}. In particular, when an X-ray photon is absorbed in the gas gap of the GPD, a photo-electron~(PE) is ejected, producing an ionization pattern that defines %\red{https://www.overleaf.com/2468253653mzhzzjkxrxjm}
a track. Each track is drifted by a uniform electric field to the \emph{Gas Electron Multiplier}~(GEM), which amplifies the charge keeping the shape unchanged~\citep{articleCosta}. The amplified charge is then read out through a grid of hexagonal pixels and the image of the PE is recorded. An example of a PE track image is given in Figure~\ref{fig:ixpetrack}: the green dot represents the impact point, where the X-ray converted into a PE, and the green line shows the emission direction of the secondary particle, which lies preferentially in the oscillation plane of the X-ray electric field. 

A correct reconstruction of the impact point is crucial to carry out the imaging of observed extended sources in the sky, while the estimation of the PE emission direction is fundamental to determine the polarization of the incoming radiation. The reconstruction of IXPE events can be reduced to the estimation of these two main parameters: (1) the impact point, and (2) the polarization angle $\phi$. Currently, an analytic approach is used to infer both the impact point and the polarization angle from the charge-waited pixel content exploiting a geometrical moments analysis technique. This approach shows its weaknesses by well reconstructing (on average) only $\sim$20$\%$ of the events, loosing mostly the low-energy ones, which are generally less featured (less elongated tracks, more spot-like). 
The weakness of the analytical approach brings up the necessity of an alternative reconstruction method and, since the IXPE track reconstruction is based on images, it is very appealing from the deep learning point of view.

Deep learning has successfully been applied in various domains such as medical image analysis, remote sensing, or astronomy~\citep{LeCunBH2015}. 
%Two trends have been the main drivers behind the shift towards such novel methods. First, massive amounts of data are nowadays available for training the corresponding models. Second, the compute power has increased dramatically as well, especially due to modern massively-parallel devices. 
In this work, we report results of two first attempts to address the estimation tasks sketched above, i.e., we propose deep neural networks for~(1) the impact point estimation and for~(2) the estimation of the emission direction. We also outline strategies for further improvements of deep learning based models for both tasks.

\section{Data and Reconstruction efficiency}
\begin{figure}[t]
\begin{center}
\includegraphics[width=0.15\textwidth]{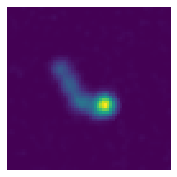}
\hfill
\includegraphics[width=0.15\textwidth]{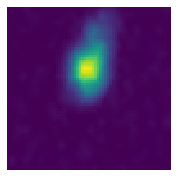}
\hfill
\includegraphics[width=0.15\textwidth]{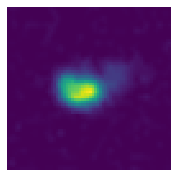}
\hfill
\includegraphics[width=0.15\textwidth]{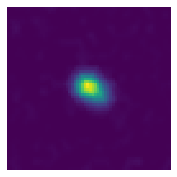}
\hfill
\includegraphics[width=0.15\textwidth]{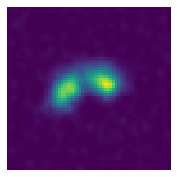}
\hfill
\includegraphics[width=0.15\textwidth]{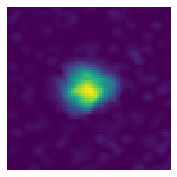}
\end{center}
\caption{Examples of simulated images}
\label{fig:examples}
\end{figure}
For this study, we simulated IXPE observations of an unpolarized source emitting X-rays of energies uniformly distributed in the range of the IXPE sensitivity. In particular, Monte Carlo simulations were used to generate 500,000 PE track images. Each track image was labelled with the following set of parameters: (1) the energy of the incoming X-ray (E$_{X}$), (2) the coordinates $(j, i)$ of the pixel containing the impact point, and (3) the polarization angle~$\phi$.
\begin{wrapfigure}{r}{10.5cm}
    \vspace{-8pt}
    \centering
    %\hskip{-10pt}
 \includegraphics[width=10.5cm]{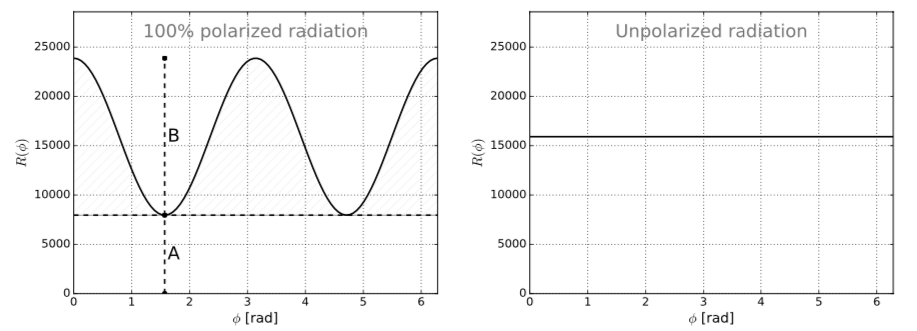}
    %\vspace{-15pt}
    \caption{Left: Illustration of a 100\% polarized radiation as seen by IXPE. Right: Same as left image but for an unpolarized radiation.}
    \label{fig:ixpetrack2}
    %\vspace{-5pt}
\end{wrapfigure}
The generated images were subsequently normalized (i.e., pixel values in [0,1]) and upsampled to a cartesian grid (upsampling factor $\approx 2$, equally-shaped square images). In Figure~\ref{fig:examples}, some examples of such generated images are shown. A separate additional test set of 35,838 observations for a $\pi/4$ polarized source emitting X-rays at 4 kilo electronvolts was processed in the same way.
% images were n
% The images of IXPE tracks as given by the simulations need to be reformatted, in order to be processed by a standard deep learning algorithm. 
% This procedure consists in a scaling of the pixels values (we want all the images to have values between 0 an 1), and a re-sampling of the image (we want equally-shaped squared images to achieve a sufficient resolution). 

A relevant characteristic is that the distribution of~$\phi$ of the events collected observing an X-ray source in the sky would show a $cos^2$ modulation in the case that the target is linearly polarized, while it would be uniform in case of unpolarized source, as illustrated in Figure \ref{fig:ixpetrack2}. The capability of an algorithm of reproducing this modulation in the final distribution of $\phi$ can be translated in an efficiency evaluation useful to compare reconstruction algorithms. We can define the modulation factor $\mu$ as the response to a fully polarized sample: the closer to 1 the modulation factor, the more efficient the algorithm. In addition to the modulation factor, we define the \textit{efficiency}, $\varepsilon_{10}$, to compare the performance of different algorithms in predicting~$\phi$. It is given by the number of events whose predicted polarization angle lies within 10 degrees
(arbitrary but reasonable number) 
of the true value. 
%\todot{[Fabian] I have removed some text here. Do we need this text actually for this work??}
% REMOVED BY FABIAN (NEEDED??)
Given the periodicity of the polarization angle distribution a shifting phase of $\pm \pi$ and $\pm$2$\pi$ of the reconstructed angle is still good, since the overall distribution would not be altered (as long as the number of positive-shifted events is balanced by the number of negative shifted events).

\section{Angle and Impact Estimation}
\label{s.exper}
We propose two different neural network architectures to estimate (a) the emission direction (angle) and (b) the impact point of a given track, respectively.
%\textcolor{red}{
%\begin{itemize}
%    \item short summary of conv nets (2-3 lines)
%   \item mention the different models used 
%    \item provide results (i.e., different plots and images, results on the test set, comparison with existing approach)
%\end{itemize}
%}
\subsection{Polarization Angle Reconstruction}
For this subtask, we resort to a M-head ensemble of VGG-16 networks with $M=8$ heads in the ensemble, where all but the last block of CNN filters are shared~\citep{mhead2015,simonyan2014}. The network produces a normalized direction vector and we consider the cosine similarity loss between the network predictions and the ground truth directions as loss function. 
Following \citep{mhead2015}, we assign a higher weight factor to the loss for the ensemble head, which gives most accurate prediction, and lower weights for the other heads in order to prevent network heads from becoming too similar. At inference stage, the average over ensemble heads is returned as the final prediction. 

\begin{figure}[t]
\begin{center}
\includegraphics[width=0.47\textwidth, height=4.5cm]{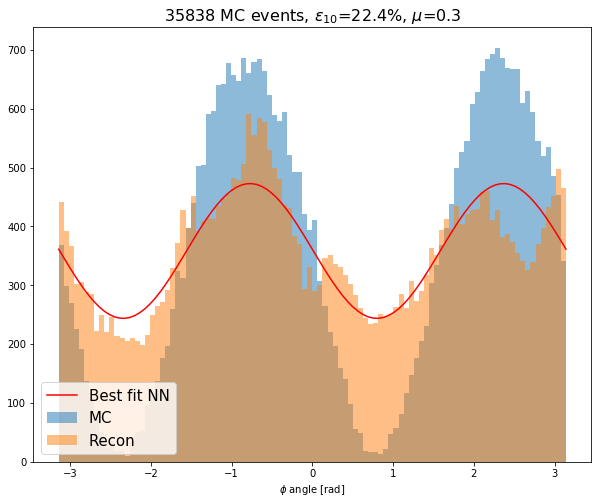}
\hfill
\includegraphics[width=0.47\textwidth, height=4.5cm]{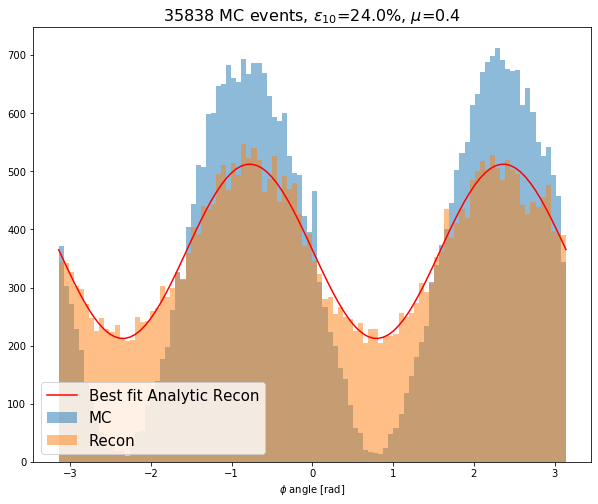}
\end{center}
\caption{Left: neural network prediction efficiency. Right: state of the art analytical method. The efficiency $\epsilon_{10}$ and the modulation factor $\mu$ are reported on top of the plots.
%Right: reconstructed impact point (marked as a red pixel on the track image).
}
\label{fig:efficiency}
\end{figure}

For typical low energy events accurate direction regression can be impossible, therefore it is important to quantify predictive uncertainty of the model, which we accomplish by using ensembling. It is known that ensembles of neural networks typically yield the best estimate of predictive uncertainty~\citep{fertig2019}, compared to methods such as dropout~\citep{gal2016} and SVI~\citep{blundel2015}, in addition to improvements in accuracy over single models. In Figure~\ref{fig:efficiency}, histograms of the reconstructions of the $\pi/4$-polarized test data  together with $10$-degree efficiency estimates $\varepsilon_{10}$ and the modulation factors $\mu$ for both the neural network and the classical reconstructions are provided. It can be seen that the baseline model gives nearly the same efficiency as the state-of-the-art analytical method. Ideally, the goal is to outperform the analytical method, and we will discuss potential strategies for that in the conclusion.

\subsection{Impact Point Reconstruction}
\begin{wrapfigure}{r}{0.5\textwidth}
\vspace{-60pt}
\begin{center}
\includegraphics[width=0.5\textwidth, height=4cm]{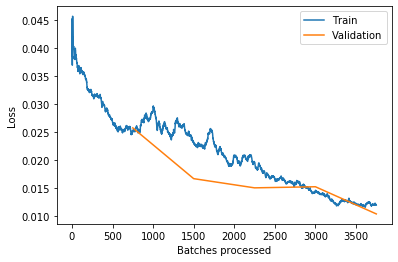}
\end{center}
%\vspace{-5pt}
\caption{Training and validation loss}
\label{fig:lossVsbatches}
\vspace{-30pt}
\end{wrapfigure}
For the impact point estimation task, we resorted to the ResNet-34 model~\citep{he2016deep} with pre-trained weights (based on ImageNet). That is, we follow a transfer learning approach~\citep{pratt1993discriminability} and only fine-tune the last layer according to the new task.
The image sample are labeled based on the true (simulated) impact point coordinates. For training the last layer, we used 60,000 events from the available track samples, 20\% samples were used for validation. We obtained 15,000 events separately for the purpose of testing the model. %(e.g. Figure 2, subfigure 3).

For training the model, we considered the Mean Square Error (MSE) as the loss function, an initial learning rate of $3e-2$ with weight decay of $1e-3$, a batch size of 64, and five epochs. The image size of the samples is $64 \times 64$. The training loss starts with a value of 0.0372 (Figure \ref{fig:lossVsbatches}) and later follows a downward path with the increase in the number of batches processed.

The validation loss starts with a similar value of 0.0257 and later decreases steadily. Near the end of the batch processing, the training and the validation loss reach a value of 0.011 and 0.010 respectively. We use Root Mean Square Error (RMSE) as the performance measure and the model showed RMSE in x to be 7.807 whereas RMSE in y as 7.368 for evaluation on test data.

\begin{figure}[t]
\begin{center}
\includegraphics[width=0.47\textwidth, height=2.5cm]{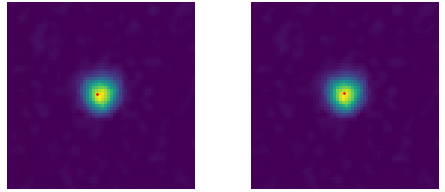}
\hfill
\includegraphics[width=0.47\textwidth, height=2.5cm]{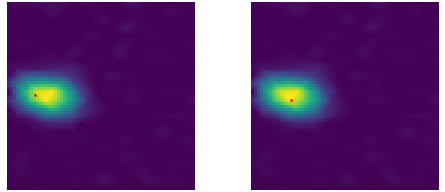}
\end{center}
\caption{Subfigure 1 and 2 - Left: Ground truth, Right: Regression prediction. 
%Right: reconstructed impact point (marked as a red pixel on the track image).
}
\label{fig:predictions}
\end{figure}

\section{Conclusions and Future Work}
\label{s.concl}
We have introduced the track reconstruction challenge for the IXPE mission and have shown that existing neural network architectures can achieve results close to the state-of-the-art reconstruction algorithms. In addition to comparable efficiency of the reconstruction, the machine learning techniques provide means of estimating the uncertainties associated to the predicted values, which is an important advantage over the analytic approach, and allows us to set quality cuts on the final reconstructed data, enhancing the accuracy of the IXPE scientific observations. Furthermore, there are multiple directions for further research:
\begin{itemize}
    \item Firstly, it is worth stressing that the current image pre-processing is not optimal: Since the original images are cropped around the cluster of pixels above the trigger threshold, they exhibit different sizes, meaning that our attempt to produce equally-sized images alters the aspect ratio of the actual tracks.
    \item Secondly, since the sensor has hexagonal pixels, conventional `cartesian' convolutional filters in fact \emph{do not} yield equivariant feature maps when applied to the raw image data coming from the sensor and can, therefore, lead to a suboptimal performance. For the baseline experiments we used $2\times$ upsampling to a cartesian grid from the original hexagonal grid, but a better approach would be to use \emph{hexagonal} convolutions instead, which work with raw data and take the actual sensor grid shape into account. Hexagonal convolutions have been implemented, e.g., in the HexagDLy library~\citep{hexagdly_paper} for PyTorch. A further step in this direction would be to investigate hexagonal group convolutions~\citep{cohen2016,hoogeboom2018}, which capture rotational feature symmetries and result in higher parameter efficiency.
   \item Thirdly, model calibration should be improved as well. We see in Figure~\ref{fig:efficiency} that, compared to the neural network, the analytical method results in a very clear sinusoidal shape of the histogram. Increasing the ensemble size and using alternative methods for sampling from the posterior distribution of directions could potentially reduce the irregularities for the neural network reconstructions. A possible improvement could be achieved by adding the information about the location of the impact point as input parameters in addition to the images.
   \item Finally, the basic direction regression and hit point detection tasks can be combined in a single model for simultaneous prediction on both tasks, as is typically done for \emph{multi-task learning} \citep{Ruder2017} tasks. Multi-task learning, intuitively, adds additional supervision signals to the network, and such additional signals could lead to an overall model outperforming the individual models trained exclusively for single tasks.
\end{itemize}
We plan to investigate the aforementioned extensions and research directions in the near future.

\newpage
\section{Acknowledgements}
We want to thank the DarkMachines collaboration for bringing us together and for fruitful discussions. Michela Negro wants to acknowledge the IXPE team and in particular Niccol\'o Di Lalla and Alberto Manfreda for providing the simulated data samples. 
\printbibliography

@ARTICLE{LeCunBH2015,
   author = {Yann LeCun and Yoshua Bengio and Geoffrey Hinton},
    title = {Deep Learning},
  journal = {Nature},
     year = {2015},
   volume = {521},
   issue = {7553},
    pages = {436-444},
   publisher = {Nature Publishing Group}
}

@ARTICLE{blundel2015,
       author = {{Blundell}, Charles and {Cornebise}, Julien and {Kavukcuoglu}, Koray and
         {Wierstra}, Daan},
        title = "{Weight Uncertainty in Neural Networks}",
      journal = {arXiv e-prints},
         year = "2015",
        month = "May",
          eid = {arXiv:1505.05424},
        pages = {arXiv:1505.05424},
archivePrefix = {arXiv},
       eprint = {1505.05424},
 primaryClass = {stat.ML},
       adsurl = {https://ui.adsabs.harvard.edu/abs/2015arXiv150505424B}
}

@ARTICLE{gal2016,
       author = {{Gal}, Yarin and {Ghahramani}, Zoubin},
        title = "{Dropout as a Bayesian Approximation: Representing Model Uncertainty in Deep Learning}",
      journal = {arXiv e-prints},
         year = "2015",
        month = "Jun",
          eid = {arXiv:1506.02142},
        pages = {arXiv:1506.02142},
archivePrefix = {arXiv},
       eprint = {1506.02142},
 primaryClass = {stat.ML},
       adsurl = {https://ui.adsabs.harvard.edu/abs/2015arXiv150602142G}
}

@article{cohen2016,
  author    = {Taco S. Cohen and
               Max Welling},
  title     = {Group Equivariant Convolutional Networks},
  journal   = {CoRR},
  volume    = {abs/1602.07576},
  year      = {2016},
  url       = {http://arxiv.org/abs/1602.07576},
  archivePrefix = {arXiv},
  eprint    = {1602.07576},
  bibsource = {dblp computer science bibliography, https://dblp.org}
}

@article{hoogeboom2018,
  author    = {Emiel Hoogeboom and
               Jorn W. T. Peters and
               Taco S. Cohen and
               Max Welling},
  title     = {HexaConv},
  journal   = {CoRR},
  volume    = {abs/1803.02108},
  year      = {2018},
  url       = {http://arxiv.org/abs/1803.02108},
  archivePrefix = {arXiv},
  eprint    = {1803.02108},
  bibsource = {dblp computer science bibliography, https://dblp.org}
}

@ARTICLE{fertig2019,
       author = {{Ovadia}, Yaniv and {Fertig}, Emily and {Ren}, Jie and {Nado}, Zachary and
         {Sculley}, D and {Nowozin}, Sebastian and {Dillon}, Joshua V. and
         {Lakshminarayanan}, Balaji and {Snoek}, Jasper},
        title = "{Can You Trust Your Model's Uncertainty? Evaluating Predictive Uncertainty Under Dataset Shift}",
      journal = {arXiv e-prints},
         year = "2019",
        month = "Jun",
          eid = {arXiv:1906.02530},
        pages = {arXiv:1906.02530},
archivePrefix = {arXiv},
       eprint = {1906.02530},
 primaryClass = {stat.ML},
       adsurl = {https://ui.adsabs.harvard.edu/abs/2019arXiv190602530O}
}

@article{hexagdly_paper,
    title = "HexagDLy—Processing hexagonally sampled data with CNNs in PyTorch",
    author = "Constantin Steppa and Tim L. Holch",
    journal = "SoftwareX",
    volume = "9",
    pages = "193 - 198",
    year = "2019",
    issn = "2352-7110",
    doi = "https://doi.org/10.1016/j.softx.2019.02.010",
    url = "https://www.sciencedirect.com/science/article/pii/S2352711018302723"
}

@article{mhead2015,
  author    = {Stefan Lee and
               Senthil Purushwalkam and
               Michael Cogswell and
               David J. Crandall and
               Dhruv Batra},
  title     = {Why {M} Heads are Better than One: Training a Diverse Ensemble of
               Deep Networks},
  journal   = {CoRR},
  volume    = {abs/1511.06314},
  year      = {2015},
  url       = {http://arxiv.org/abs/1511.06314},
  archivePrefix = {arXiv},
  eprint    = {1511.06314},
  bibsource = {dblp computer science bibliography, https://dblp.org}
}

@article{articleCosta,
author = {Costa, Enrico and Soffitta, Paolo and Bellazzini, Ronaldo and Brez, Alessandro and Lumb, Nicholas and Spandre, Gloria},
year = {2001},
month = {07},
pages = {662-5},
title = {An efficient photoelectric X-ray Polarimeter for the study of Black Holes and Neutron Stars},
volume = {411},
journal = {Nature},
doi = {10.1038/35079508}
}

@ARTICLE{2006NIMPA.560..425B,
       author = {{Bellazzini}, R. and {Angelini}, F. and {Baldini}, L. and {Bitti}, F. and
         {Brez}, A. and {Cavalca}, F. and {Del Prete}, M. and {Kuss}, M. and
         {Latronico}, L. and {Omodei}, N.},
        title = "{Gas pixel detectors for X-ray polarimetry applications}",
      journal = {Nuclear Instruments and Methods in Physics Research A},
         year = "2006",
        month = "May",
       volume = {560},
       number = {2},
        pages = {425-434},
          doi = {10.1016/j.nima.2006.01.046},
archivePrefix = {arXiv},
       eprint = {astro-ph/0512242},
 primaryClass = {astro-ph},
       adsurl = {https://ui.adsabs.harvard.edu/abs/2006NIMPA.560..425B}
}

@Article{galaxies6010033,
AUTHOR = {Weisskopf, Martin C.},
TITLE = {An Overview of X-Ray Polarimetry of Astronomical Sources},
JOURNAL = {Galaxies},
VOLUME = {6},
YEAR = {2018},
NUMBER = {1},
ARTICLE-NUMBER = {33},
URL = {https://www.mdpi.com/2075-4434/6/1/33},
ISSN = {2075-4434},
DOI = {10.3390/galaxies6010033}
}

@ARTICLE{simonyan2014,
       author = {{Simonyan}, Karen and {Zisserman}, Andrew},
        title = "{Very Deep Convolutional Networks for Large-Scale Image Recognition}",
      journal = {arXiv e-prints},
         year = "2014",
        month = "Sep",
          eid = {arXiv:1409.1556},
        pages = {arXiv:1409.1556},
archivePrefix = {arXiv},
       eprint = {1409.1556},
 primaryClass = {cs.CV},
       adsurl = {https://ui.adsabs.harvard.edu/abs/2014arXiv1409.1556S}
}

@article{Ruder2017,
  author    = {Sebastian Ruder},
  title     = {An Overview of Multi-Task Learning in Deep Neural Networks},
  journal   = {CoRR},
  volume    = {abs/1706.05098},
  year      = {2017},
  url       = {http://arxiv.org/abs/1706.05098},
  archivePrefix = {arXiv},
  eprint    = {1706.05098},
  bibsource = {dblp computer science bibliography, https://dblp.org}
}

@inproceedings{pratt1993discriminability,
  title={Discriminability-based transfer between neural networks},
  author={Pratt, Lorien Y},
  booktitle={Advances in neural information processing systems},
  pages={204--211},
  year={1993}
}

@inproceedings{he2016deep,
  title={Deep residual learning for image recognition},
  author={He, Kaiming and Zhang, Xiangyu and Ren, Shaoqing and Sun, Jian},
  booktitle={Proceedings of the IEEE conference on computer vision and pattern recognition},
  pages={770--778},
  year={2016}
}
\end{document}